\documentstyle[11pt,a4]{article}

\newcommand{\be}{\begin{equation}}
\newcommand{\ee}{\end{equation}} 
\newcommand{\bea}{\begin{eqnarray}}
\newcommand{\eea}{\end{eqnarray}}

\catcode`\@=11 
\@addtoreset{equation}{section}
 
\catcode`\@=11

\begin{document}

\begin{titlepage}

\begin{flushright} 
{\tt 	FTUV/97-10\\ 
	IFIC/97-10\\ }
 \end{flushright}

\bigskip

\begin{center}

{\bf \LARGE Free Fields via Canonical Transformations of \\ 
	Matter-coupled 2D Dilaton Gravity Models}
\footnote{Work partially supported by the 
{\it Comisi\'on Interministerial de Ciencia y Tecnolog\'{\i}a}\/ 
and {\it DGICYT} (under contracts AEN96-1669 and PB95-1201).}

\bigskip 

 J.~Cruz$^a$\footnote{\sc cruz@lie.uv.es},
 J.  M.  Izquierdo$^b$\footnote{\sc izquierd@lie.uv.es},  
 D.  J.  Navarro$^a$\footnote{\sc dnavarro@lie.uv.es} and
 J.~Navarro-Salas$^a$\footnote{\sc jnavarro@lie.uv.es}.

\end{center}

\bigskip%

\footnotesize
	
a) Departamento de F\'{\i}sica Te\'orica and 
	IFIC, Centro Mixto Universidad de Valencia-CSIC.
	Facultad de F\'{\i}sica, Universidad de Valencia,	
        Burjassot-46100, Valencia, Spain. 
   \newline               

b) Departamento de F\'{\i}sica Te\'{o}rica. Facultad de Ciencias, Universidad de Valladolid,
Valladolid, 47011-Valladolid, Spain.
\normalsize 

\bigskip
\bigskip


\begin{center}
			{\bf Abstract}
\end{center}
It is shown that the 1+1-dimensional matter-coupled Jackiw-Teitelboim model 
and the model with an exponential potential
 can be converted by means of 
 appropriate canonical transformations into a bosonic
 string theory propagating on a flat target space with an indefinite signature.
 This makes it possible to consistently 
 quantize these models
 in the functional Schr\"odinger representation
 thus generalizing recent results on the CGHS theory.
 \newline
 \newline
 PACS number(s): 04.60.Kz, 04.60.Ds
 \newline
 \end{titlepage}
 \newpage

 \section{Introduction}
 In recent years there has been a lot of interest in two-dimensional
 dilaton-gravity theories. The main motivation is that they posses most of the 
 interesting physical features of the four-dimensional theory such as the formation of black holes
 and their subsequent evaporation, while the technical difficulties are reduced.
 The simplest theory describing the formation and evaporation of 2D
 black holes is the model introduced by Callan, Giddings, Harvey
 and Strominger (CGHS) \cite{CGHS}.
 The semiclassical approach as well as the canonical quantization of the model
 have been considered from different viewpoints \cite{Strominger}.
 In a recent work \cite{Cangemi} it was shown that the CGHS
 model can be converted, by means of a canonical
 transformation, into a bosonic string theory propagating
 on a Minkowskian target space.
 A further canonical transformation brings the constraints into the form of a 
 parametrized theory \cite{Kuchar1}.
 At the quantum level the commutators of the constraint
 operators produce the Virasoro anomaly and the theory cannot be
 quantized without modification.
However the value of the anomaly depends on the choice of the
 vacuum used to normal order the constraints.
  The one associated with the Schr\"odinger representation
 produces a cancellation between the anomaly of the two free fields of the 
 pure gravity theory \cite{Cangemi}.
 Therefore it is possible to find solutions to the quantum
 Dirac constraint algebra \cite{Louis,Cangemi}.
 For the matter-coupled theory a consistent quantum theory can be
 constructed on a cylinder by an appropriate 
 modification of the quantum constraints 
 which removes the anomaly \cite{Benedict}.
 This mechanism is based on an embedding-dependent factor ordering
 of the constraints \cite{Kuchar1}.
 The extension of the anomaly-free Dirac quantization
 of matter-coupled CGHS theory
 to an open two-dimensional space-time has been given in \cite{Kuchar2}
  (see also \cite{Louis2}).
 This quantization procedure yields to a physical spectrum in accordance 
 with the degrees of freedom of the classical theory
 \cite{Benedict,Kuchar2}.
 The BRST quantization gives an inequivalent physical result \cite{Benedict}.
The aim of this paper is to generalize the results of \cite{Benedict,Kuchar2} 
concerning the CGHS model to other models of two-dimensional dilaton gravity.

Up to conformal redefinitions of the fields the action of a generic
(first-order) model of
2D dilaton gravity can be written in the form
 \cite{Louis}
 \be
 S=\int d^2x\sqrt{-g}\left[R\phi+V\left(\phi\right)
 -{1\over2}\left(\nabla f\right)^2\right]\>.\label{ai}
 \ee
 where $V\left(\phi\right)$ is an arbitrary potential term.
 The CGHS model can be recovered by choosing a constant potential $V\left(\phi\right)=4\lambda^2$.
 If we parametrize the two-dimensional metric as
   \be
  g_{\mu\nu}=e^{2\rho}\left(\begin{array}{cc}v^2-u^2&v\\v&1\end{array}\right)\>,\label{aii}
  \ee
  where the functions $u$ and $v$ are related to the shift and lapse functions,
  the hamiltonian form of the action becomes
   \be
  S=\int d^2x\left(\pi_{\rho}\dot\rho+\pi_{\phi}\dot\phi+\pi_f\dot f-uH-vP\right)\>,\label{aiii}
  \ee 
    where the constraint functions are given by
 \be
 H=-{1\over2}\pi_{\rho}\pi_{\phi}+2\left(\phi^{\prime\prime}-\phi^{\prime}\rho^{\prime}
 \right)-e^{2\rho}V\left(\phi\right)+{1\over2} 
 \left(\pi_{f}^2+f^{\prime 2}\right)\>,\label{aiv}
 \ee
 \be
 P=\rho^{\prime}\pi_{\rho}-\pi_{\rho}^{\prime}
 +\phi^{\prime}\pi_{\phi}+\pi_{f}f^{\prime}\>.\label{av}
 \ee
 In \cite{Cangemi} it has been shown that when the potential $V\left(\phi\right)$ is constant,
a canonical transformation converts the constraints $H$ and $P$ 
into those a bosonic string theory propagating on a 3-dimensional Minkowski space
\be
H={1\over2}\left(\pi^2_0+\left(r^{0\prime}\right)^2 \right)-{1\over2}\left(\pi^2_1+\left(r^{1\prime}\right)^2\right)
+{1\over2}\left(\pi_f^2+f^{\prime 2}\right)
\>,\label{avi}
\ee
\be
P=\pi_0r^{0\prime}+\pi_1r^{1\prime}+\pi_ff^{\prime}
\>.\label{avii}
\ee

In this paper we shall show that the Jackiw-Teitelboim model $\left(V\left(\phi\right)=4\lambda^2\phi\right)$
and the model with an exponential potential $\left(V\left(\phi\right)=4\lambda^2e^{\beta\phi}\right)$,
which includes the CGHS model as a limiting case $\left(\beta =0\right)$, can also be converted 
through a canonical transformation into a bosonic string theory propagating on a Minkows\-kian 
target space.
In conformal gauge the exponential model possesses a free field $\eta$ and a Liouville one $\varphi$.
It is well known that a Liouville field $\varphi$ can be mapped into a 
free field $\psi$ through a canonical transformation.
However the energy momentum tensor of the two free fields $\eta$ and $\psi$ contains an 
"improvement"
term which does not allow to implement the Schr\"odinger quantization approach
\cite{Benedict}.
In fact the improvement terms are naturally related to the
BRST quantization scheme.
Therefore a further canonical transformation
mixing the two free fields is required to transform the exponential model into a bosonic string theory
with a Minkowskian target-space.
In section 2 we shall construct this canonical transformation
directly from the $\rho$ and $\phi$ fields.
As a by-product we shall also show how to
recover the canonical transformation of the CGHS theory \cite{Kuchar2} as a limiting case $\beta=0$
of our approach.
Moreover we shall carry out the Dirac constraint quantization of the model
following the lines of \cite{Benedict}.
In section 3 we shall analyze the Jackiw-Teitelboim model.
This model is more involved because it is described by a Liouville field and a field
propagating in a De Sitter space with a curvature term.
Both fields can be also combined, through a canonical transformation, to
produce two free fields without any improvement term and with opposite
contributions to the hamiltonian constraint.
Therefore this theory is also equivalent
 to a bosonic string theory with a Minkowskian
  target space. These results make it possible to consistently quantize
  these theories.

 \section{The Exponential Model}
 \subsection{ Canonical Transformation}
 Lets us consider the model (\ref{ai})
 with $V\left(\phi\right)=4\lambda^2 e^{\beta\phi}$. 
 This model includes the CGHS theory as a particular case
 $\beta=0$. In the absence of matter fields the above model possesses static
 black hole solutions similar to the ones of the CGHS model,
 but in contrast with them
 these black holes have a Hawking temperature 
  proportional to their
  mass \cite{Mann}.
  Due to the existence of an extra symmetry,
  it is possible, in analogy with the CGHS theory,
  to construct a solvable semiclassical theory \cite{Cruz1,Cruz2}.
  The semiclassical analysis indicates that these black holes never
  disappear completely. 
   The equations of motion of the model, in conformal gauge ($ds^2=
   -e^{-2\rho}dx^+dx^-$),
  are
  \be
  \partial_+\partial_-\left(2\rho-\beta\phi\right)=0
  \>, \label{bii}
  \ee
  \be
  \partial_+\partial_-\left(2\rho+\beta\phi\right)=-2\lambda^2\beta e^{2\rho+\beta\phi}
  \>, \label{biii}
  \ee
  \be
  \partial_{\pm}^2\phi-2\partial_{\pm}\rho\partial_{\pm}\phi=T^f_{\pm\pm}={1\over2}\left(\partial_{\pm}f\right)^2
  \>,\label{biv}
  \ee
  \be
  \partial_+\partial_-f=0
  \>.\label{bv}
  \ee
   The unconstrained equations (\ref{bii}-\ref{biii}) are equivalent to a free field 
  equation and a Liouville equation respectively. The general solution to these 
  equations suggests the following transformation
   in terms of a set of new variables
  $A_{\pm},a_{\pm}$ 
\be
 \rho={1\over2}\log {-A^{\prime}_+A^{\prime}_-\over 1+\lambda^2\beta A_+A_-}
 -{\beta\over2}\left( a_+ +a_-\right)\>,\label{bviii}
 \ee
  
  \be
  \pi_{\rho}=2\lambda^2{\left(A_+^{\prime}A_--A_+A_-^{\prime}\right)\over 1+\lambda^2\beta A_+A_-}
  -2\left(a_+^{\prime}- a_-^{\prime}\right)\>, \label{bix}
  \ee
 
 \be
 \phi=-{1\over\beta}\log\left(1+\lambda^2\beta A_+A_-\right)+ a_++a_-\>,\label{bx}
  \ee

  \be
  \pi_{\phi}=-\left({A^{\prime\prime}_+\over A_+^{\prime}}
  -{A_-^{\prime\prime}\over A_-^{\prime}}\right)
  +\lambda^2\beta {\left(A_+^{\prime}A_--A_+A_-^{\prime}\right)
  \over 1+\lambda^2\beta A_+A_-}+\beta\left(
   a_+^{\prime}- a_-^{\prime}\right)
  \>.\label{bxi}
  \ee
  The canonical structure of the theory can be equivalently described
  by the 2-form $\omega$ defined by
  \be
  \omega=\int dx\left(\delta\rho\wedge\delta\pi_{\rho}
  +\delta\phi\wedge\delta\pi_{\phi}+\delta f\wedge\delta
  \pi_f\right)\>.\label{bxiv}
  \ee
   In terms of the fields $A_{\pm},a_{\pm}$ this 2-form turns out to be
   \bea
   \omega&=&\int dx\left[\delta a_+\wedge \delta\left(-2{A_+^{\prime\prime}\over 
   A_+^{\prime}}
   +2\beta a_+^{\prime}
   +2{a_+^{\prime\prime}\over a_+^{\prime}}\right)\right.+\nonumber\\
   &&\left. \delta a_-\wedge\delta\left(2{A_-^{\prime\prime}\over A_-^{\prime}}-2\beta a_-^{\prime}
   -2{a_-^{\prime\prime}\over a_-^{\prime}}\right)+\delta f\wedge\delta\pi_f\right]
  +\omega_b\>,\label{bxv}
  \eea
  where $\omega_b$ is just a boundary term
  (from now on the exterior product will be omitted)
   \bea
   \omega_b&= &\int d \left[-{\delta A_+^{\prime}\over A_+^{\prime}}\delta a_+
   +{\delta A_-\over A_-^{\prime}}\delta a_-+\delta a_+{\delta A_-^{\prime}\over A_-^{\prime}} 
   -\delta a_-{\delta A_+^{\prime}\over A_+^{\prime}}\right.\nonumber\\
   && -2\beta\delta a_+\delta a_-+2{\delta a_+^{\prime}\over a_+^{\prime}}
   \delta a_+-2{\delta a_-^{\prime}\over a_-^{\prime}}\delta a_-\nonumber\\
   && +\lambda^2\left({A_-\over A_+^{\prime}}\delta A_+\delta A_+^{\prime}
   -{A_+ \over A_-^{\prime}}\delta A_-\delta A_-^{\prime}+{A_+\over A_+^{\prime}}\delta A_-\delta A_+^{\prime}
   +{A_- \over A_-^{\prime}}\delta A_+ \delta A_-^{\prime}\right.\nonumber\\
 &&  +2\delta A_+\delta A_--{1\over2}{\delta\left(A_+A_-\right)\over 1+\lambda^2\beta A_+A_-}
   \left({\delta A_+^{\prime}\over A_+^{\prime}}-{\delta A_-^{\prime}\over A_-^{\prime}}\right)
  \nonumber\\  && \left. \left. -{\delta A_+\delta A_-\over 1+\lambda^2\beta A_+A_-}\right)\right]
   \>, \label{bxvi}
   \eea
   and the constraints $C_{\pm}=\pm {1\over2}\left(H\pm P\right)$ become
   \be
   C_{\pm}=\mp 2 a_{\pm}^{\prime}\left({A_{\pm}^{\prime\prime}\over A_{\pm}
  ^{\prime}}-\beta a_{\pm}^{\prime}-{a_{\pm}^{\prime\prime}
 \over a_{\pm}^{\prime}}\right)\pm {1\over4}\left(\pi_f\pm f^{\prime}\right)^2\>.\label{bxvii}
 \ee
 At this point it is clear that the following additional transformation
 $\left(A_{\pm},a_{\pm}\right)\rightarrow\left(X^{\pm},\Pi_{\pm}\right)$
 \be
 X^{\pm}=a_{\pm}\>,\label{bxviii}
  \ee
  \be
  \Pi_{\pm}=\mp 2
  \left({A_{\pm}^{\prime\prime}\over A_{\pm}
  ^{\prime}}-\beta a_{\pm}^{\prime}-{a_{\pm}^{\prime\prime}
 \over a_{\pm}^{\prime}}\right)
 \>.\label{bxix}
 \ee
 implies that
 \be
 \omega=\int dx \left(\delta X^+\delta \Pi_+ 
 +\delta X^-\delta \Pi_-+\delta f\delta \pi_f \right)+\omega_b  \>,  \label{bxx}
  \ee
  and brings the constraints into the form of a parametrized scalar field theory
  on a flat background \cite{Kuchar3}
  \be
  C_{\pm}=\Pi_{\pm}X^{\pm\prime}\pm {1\over4}\left(\pi_f\pm f^{\prime}\right)^2
  \>,\label{bxxi}
  \ee
  The composition of (\ref{bviii}-\ref{bxi}) with the inverse of
  (\ref{bxviii}-\ref{bxix})
  \be
  a_{\pm}=X^{\pm}\>,\label{bxxii}
  \ee
  \be
  A_{\pm}=\pm \int^x \exp \int^x \mp {1\over2} \Pi_{\pm}+\beta 
  X^{\pm\prime}+\left(\log X^{\pm\prime}\right)^{\prime} 
  \>,\label{bxxiii}
  \ee
  defines, up to the boundary term, a canonical transformation.
  However if we restrict the analysis to a closed spatial section $\left(x\in \left[
  0,2\pi\right]\right)$
  the boundary contribution to the 2-form $\omega$
  \be
  \omega_b=-2\delta\left(\log {A_+^{\prime}A_-^{\prime}\over X^{+\prime}
  X{^\prime}}-\beta\left(X^++X^-\right)\right)\left(0\right)
  \delta\left(X^+\left(2\pi\right)-X^+\left(0\right)\right) \>, \label{bxxiv}
  \ee
  vanishes fixing the monodromy of the fields $X^{\pm}$ as follows (in a parallel way to the CGHS theory
  \cite{Benedict})
  \be
  X^{\pm}\left(2\pi \right)-X^{\pm}\left(0\right)=\pm 2\pi \>. \label{bxxv}
  \ee
  These conditions are consistent with the requirement
  $X^{\pm\prime}\neq 0$ needed to have a non-singular transformation.
  A further transformation in the gravitational sector
   \cite{Kuchar1,Cangemi,Benedict}
   \be
   2\Pi_{\pm}=-\left(\pi_0+\pi_1\right)\mp \left(r^{0\prime}-r^{1\prime}\right)
   \>,\label{bxxvi}\ee
   \be
   2X^{\pm\prime}=\mp\left(\pi_0-\pi_1\right)-\left(r^{0\prime}+r^{1\prime}\right)\>,\label{bxxvii}
   \ee
   casts the constraints of the matter-coupled gravity theory into
   those of a bosonic string in a 3-dimensional Minkowskian
   target-space (\ref{avi}-\ref{avii}).
   
  At this point it is interesting to consider the case $\beta=0$ (i.e, the CGHS model).
  When $\beta=0$ we can alternatively rewrite the two-form
  $\omega$ as 
  \be
  \omega=\int dx\left[-\delta A_+\delta\left({a_+^{\prime}\over A_+^{\prime}}\right)^{\prime}
  +\delta A_-\delta \left({a_-^{\prime}\over A_-^{\prime}}\right)^{\prime}+
  \delta f\delta\pi_f\right]+\tilde\omega_b
  \>,\label{bxxviii}
  \ee 
   where $\tilde\omega_b$ is a new boundary term and a factor
   -2 has been absorbed in the arbitrary functions $a_{\pm}$.
    Defining now the canonical variables as
   \be
   X^{\pm}={\mp}A_{\pm}
   \>,\label{bxxix}
   \ee
   \be
   \Pi_{\pm}=\left({a_{\pm}^{\prime}\over A_{\pm}^{\prime}}\right)^{\prime}
   \>,\label{bxxx}
   \ee
   the constraints take the standard form (\ref{bxxi}). Composing now
  (\ref{bviii}-\ref{bxi}) with the inverse of (\ref{bxxix}-\ref{bxxx})
   we recover immediately the canonical transformation
  proposed in \cite{Kuchar2}. In the general case $\left(\beta\neq 0\right)$ it is no longer possible
  to choose the embedding fields $A_{\pm}$ as a commuting set of canonical variables and the
  natural choice is (\ref{bxviii}-\ref{bxix}). 
  
 Having found the canonical transformation, we are now exactly in the situation
described in \cite{Benedict}, so it is possible to perform the same analysis from
this point on, and we shall sketch it here for completeness.
  At the quantum level the constraints (\ref{bxxi}) close down an anomalous algebra
  \bea
  &\left[C_{\pm}\left(x\right),C_{\pm}\left(\tilde x\right)\right]=
  i\left(C_{\pm}\left(x\right)+C_{\pm}\left(\tilde x\right)\pm{1\over24\pi}
  \right)\delta^{\prime}
  \left(x-\tilde x
\right)\nonumber \\
&\mp{i\over24\pi}\delta^{\prime\prime\prime}\left(x-\tilde x\right)
\>,&\label{bxxxi}
\eea
\be
\left[C_+\left(x\right),C_-\left(\tilde x\right)\right]=0
\>,\label{bxxxii}
\ee
and the theory can not be quantized without modification.
Remarkably, the addition of a term depending on the coordinate fields
$X^{\pm}$ \cite{Cangemi,Kuchar1} cancels the anomaly
and the new constraints
\be
C_{\pm}\left(x\right)\pm{1\over48\pi}\left[\log\pm X^{\pm\prime}\right]^{\prime\prime}
\mp {1\over48\pi} \>,\label{bxxxiii}
 \ee
 satisfy the algebra (\ref{bxxxi}-\ref{bxxxii}) without centre.
 To solve the new Dirac quantization condition one can construct a quantum canonical
 transformation based on an expansion of the matter fields in terms of
 "gravitationally dressed" mode operators
 \be
 a_n^{\pm}\equiv {1\over2\sqrt{\pi}}\int^{2\pi}_0 d x\ e^{inX^{\pm}}\left(\pi_f\pm f^{\prime}
 \right) \>.\label{bxxxiv}
 \ee
 If we order the fields with respect to the mode operators (\ref{bxxxiv})
 the constraint algebra is modified but a new modification of the 
 constraints
 \be
 \bar C_{\pm}=C_{\pm}\mp{1\over48\pi}X^{\pm\prime}\left[X_{\pm}^{\prime}+\left({1\over X^{\pm\prime}}\right)^{\prime
 \prime}\right]
 \>,\label{bxxxv}
 \ee
 leads the algebra (\ref{bxxxi}-\ref{bxxxii}) without the central terms.
 Moreover the transformation
 \be
 \bar X^{\pm \prime}\bar\Pi_{\pm}=\bar C_{\pm}
 \>,\label{bxxxvi}
 \ee
 \be
 \bar X^{\pm}=X^{\pm}
 \>,\label{bxxxvii}
 \ee
 is a quantum canonical transformation and brings the quantum constraints to the simple form (\ref{bxxi}).
 The physical states are constructed by acting with the creation operators
 $a^{\pm}_{-|n|}$ on the zero-mode states $|p>$ defined by
 \be
 a^-_0|p>\equiv a_0^+|p>=p|p>
 \>,\label{bxxxviii}
 \ee
 \be
 a_n^{\pm}|p>=0\ \ n>0
 \>,\label{bxxxix}	  
  \ee
  and verify the level-matching condition as in the CGHS theory \cite{Benedict}.
  This restriction comes from the integral condition
  \be
  \int_0^{2\pi} d x\left(\Pi_+-\Pi_-\right)=0
  \>,\label{bxl}
  \ee
as can be seen immediately from inspection of (\ref{bxix}) and (\ref{bxxv}). 

\subsection{Relation with Liouville theory}

To finish this section we would like to discuss the relation of our approach to 
the 
standard one of Liouville theory. In terms of a new metric ${\tilde g}_{\mu
\nu}=e\sp{-\beta\phi}g_{\mu\nu}$ the action (\ref{ai}) in the absence of matter 
fields takes the form
\begin{equation}
          \int d\sp 2 x\sqrt{-{\tilde g}}({\tilde R}\phi+\beta \left(
{\tilde \nabla}\phi\right)\sp 2+ 4\lambda\sp 2 e\sp{2\beta\phi})\quad . \label{li}
\end{equation}
The Liouville lagrangian can be recovered from (\ref{li}) by making use of 
one of the generally covariant equations of motion, ${\tilde R}=0$, and 
fixing the gauge with the choice of the flat metric $d{\tilde s}\sp 2=-dx\sp + dx\sp -$. The
resulting theory has been shown to be canonically equivalent to a free field 
one (see, for instance \cite{BCT,HJ,GN}).
We shall now explain why this result does not give directly the above 
discussed string theory formulation of the exponential model, as might appear 
to be the case because this model is described in the conformal gauge by a 
Liouville field $\varphi= 2\rho+\beta\phi$ and a free one $\eta=2{\tilde 
\rho}= 2\rho-\beta\phi$, (see (\ref{bii}), (\ref{biii})) with certain constraints.
To this end let us first write $H$ and $P$ (\ref{aiv}-\ref{av}) in terms of
$\varphi$ and $\eta$ (we omit the matter fields $f$ for simplicity)
\begin{eqnarray}
H&=&-(\beta\pi_\varphi\sp 2 +\frac{1}{4\beta}\varphi'\sp 2-4\lambda\sp 
2e\sp\varphi-\frac{1}{\beta}\varphi'')+\beta\pi_\eta\sp 
2+\frac{1}{4\beta}\eta'\sp 2-\frac{1}{\beta}\eta'' \>,\label{lii}\\
P&=&\pi_\varphi \varphi'+\pi_\eta\eta'-2\pi_\varphi'-2\pi_\eta' \quad.
                                                      \label{liii}
\end{eqnarray}
Note that neither the Liouville nor the free field parts of $H$ and $P$ 
correspond to the canonical energy momentum tensor due to the presence of the 
spatial derivative terms $\varphi'',\eta'',\pi_\varphi',\pi_\eta'$. Rather, 
they correspond to the improved one, which can be obtained (as shown in
\cite{RJ}) by varying the 
metric $\tilde g$ in (\ref{li}) (or, in terms of $H$ and $P$, by simply 
substituting $\varphi$ and $\eta$ for $\phi$ and $\rho$ in (\ref{aiv}) and
(\ref{av})). A canonical transformation that relates the Liouville field to
 a 
free one is the following (see also \cite{HJ})
\begin{eqnarray}
\varphi&=&\psi-2\log (1+\lambda\sp 2\beta A_+ A_-)\>,\label{liv}\\
\pi_\varphi&=&\pi_\psi -\lambda\sp 2\frac{A_+'A_--A_+A_-'}{(1+\lambda\sp 
2\beta A_+A_-)} \quad ,
\label{lv}
\end{eqnarray}
where
\begin{equation}
A_+=\int\sp x\exp \int\sp x (\frac{\psi'}{2}+\beta \pi_\psi)\quad ,\quad
A_-=-\int\sp x\exp \int\sp x(\frac{\psi'}{2}-\beta\pi_\psi)\quad ,
\label{lvi}
\end{equation}
and the boundary term vanishes as can be checked by writing $\psi,\pi_\psi$
in terms of $X\sp \pm,\pi_\pm$ and then using (\ref{bxxv}).
The constraints are given in terms of the new variables as
\begin{eqnarray}
H&=&-(\beta\pi_\psi\sp 2+\frac{1}{4\beta}\psi'\sp 
2-\frac{1}{\beta}\psi'')+\beta\pi_\eta\sp 2+\frac{1}{4\beta}\eta'\sp 
2-\frac{1}{\beta}\eta''\quad ,\label{lvii}\\
P&=&\pi_\psi \psi' + \pi_\eta\eta' -2\pi_\psi'-2\pi_\eta'\quad , 
\label{lviii}
\end{eqnarray}
which correspond to the difference of two free {\it improved} energy momentum 
tensors.
>From the point of view of Liouville theory this is 
enough. However, the goal here was to connect the generally covariant theory 
given by (\ref{ai}) with string theory in order to examine its Schr\"odinger 
quantization. This demands that the improvement terms must disappear due to the 
canonical transformation, which is not the case here. Earlier we were able to 
obtain such a transformation, so it is immediate to write down a new one that 
removes the improvement terms from the $\psi$ and $\eta$ pieces of $H$ and $P$ 
in (\ref{lvii}-\ref{lviii})
\be
\psi'=\frac{1}{2}(r\sp 0{'}-r\sp 1{'})-\beta (r\sp 0{'}+r\sp 1{'})+
\left(\log[(r\sp 0+r\sp 1)'\sp 2-(\pi_0
-\pi_1)\sp 2]\right)\sp \prime \>,\label{lix}
\ee
\be
\pi_\psi =-\frac{1}{2}(\pi_0-\pi_1)+\frac{1}{4\beta}(\pi_0+\pi_1)
+\frac{1}{2\beta}\left(\log\left[\frac{(r\sp 0+r\sp 
1)'+(\pi_0-\pi_1)}{(r\sp 0+r\sp 1)'-(\pi_0-\pi_1)}\right]
\right)\sp \prime \>,\label{lx}
\ee
\be
\eta'=\frac{1}{2}(r\sp 0{'}-r\sp 1{'})+\beta (r\sp 0{'}+r\sp 1{'})+
\left(\log[(r\sp 0+r\sp 1)'\sp 2-(\pi_0
-\pi_1)\sp 2]\right)\sp \prime \>,\label{lxi}
\ee
\be
\pi_\eta =-\frac{1}{2}(\pi_0-\pi_1)-\frac{1}{4\beta}(\pi_0+\pi_1)
-\frac{1}{2\beta}\left(\log\left[\frac{(r\sp 0+r\sp
1)'+(\pi_0-\pi_1)}{(r\sp 0+r\sp 1)'-(\pi_0-\pi_1)}\right]
\right)\sp \prime  \quad .
\label{lxii}
\ee
After this transformation $H$ reads
\begin{equation}
 H=\frac{1}{2}(\pi_0\sp 2+(r\sp 0{'})\sp 2)-\frac{1}{2}(\pi_1\sp 2+
(r\sp 1{'})\sp 2)
\quad .\label{lxiii}
\end{equation}
Note that this transformation mixes the $\psi$ and $\eta$ fields up, so it 
cannot be used to achieve the same for $\psi$ alone.

   \section{The Jackiw-Teitelboim model}
The Jackiw-Teitelboim model \cite{Jackiw}
coupled to conformal matter is given by the action
\be
S= \int d^2x\sqrt{-g}\left[R\phi+4\lambda^2 \phi-
{1\over2}\left(\nabla f\right)^2\right]
\>, \label{ci}
\ee 
and it is also one of the most relevant models of 2D dilaton-gravity.
The equations of motion, in conformal gauge, become
\be
2e^{-2\rho}\partial_+\partial_-\rho+\lambda^2=0
\>,\label{cii}
\ee
\be
e^{-2\rho}\partial_+\partial_-\phi+\lambda^2\phi=0
\>,\label{ciii}
\ee
plus the constrained and matter equations (\ref{biv}-\ref{bv}), which are model independent.
Equation (\ref{cii}) implies that $\rho$ is a Liouville field so
the general solution for $\rho$ is
\be
\rho={1\over2}\log {\partial_+A_+\partial_-A_-\over 
\left(1+{\lambda^2\over2}A_+A_-\right)^2}
\>,\label{civ}
\ee
where $A_{\pm}$ is an arbitrary function depending on the $x^{\pm}$ coordinate.
Taking into account (\ref{civ}) and the constrained equations
 it is possible to find the following solution to the equation
(\ref{ciii})
\be
\phi=-{1\over2}\left({\partial_+a_+\over \partial_+A_+}
-{\partial_-a_-\over \partial_-A_-}
\right)+{\lambda^2\over2}
{\left( a_+A_--a_-A_+\right)\over 1+{\lambda^2\over2}A_+A_-}
\>.\label{cv}
\ee
with $a_{\pm}$ an arbitrary function of the $x^{\pm}$ coordinate.
Therefore the general solution is parametrized by four arbitrary
chiral functions.
Two of them are simply the two gauge fixing functions associated with conformal coordinate
transformations and the other two 
account for the two chiral sectors of the matter field.
The general solution (\ref{civ}-\ref{cv}) suggests 
the following transformation to the new variables $A_{\pm},a_{\pm}$   
( from now on $A_{\pm}$ and $a_{\pm}$ are not required to be chiral functions)

\be
\rho={1\over2}\log {-A_+^{\prime}A_-^{\prime}\over \left(1+{\lambda^2\over2}A_+A_-\right)^2}
\>,\label{cvi}
\ee
\bea
\pi_{\rho}&=&\left({a_+^{\prime}\over A_+^{\prime}}\right)^{\prime}
+\left({a_-^{\prime}\over A_-^{\prime}}\right)^{\prime}+{\lambda^4\over2}
{\left(A_+^{\prime}A_--A_+A_-^{\prime}\right)\over \left(1+{\lambda^2\over2}A_+A_-\right)^2}
\left(a_+A_--a_-A_+\right)\nonumber\\
&&-\lambda^2 {\left(a_+A_--a_-A_+\right)^{\prime} \over 1+{\lambda^2\over2}A_+A_-}
\>,\label{cvii}
\eea
\be
\phi=-{1\over2}\left({a_+^{\prime}\over A_+^{\prime}}-{a_-^{\prime}\over A_-^{\prime}}\right)
+{\lambda^2\over2}{ \left(a_+A_--a_-A_+\right) \over 1+{\lambda^2\over2}A_+A_-}
\>,\label{cviii}
\ee
\be
\pi_{\phi}=-\left({A_+^{\prime\prime}\over A_+^{\prime}}-{A_-^{\prime\prime}\over A_-^{\prime}}
\right)-2\lambda^2{\left(A_+^{\prime}A_--A_+A_-^{\prime}\right)\over 1+{\lambda^2
\over2}A_+A_-}
\>,\label{cix}
\ee
The 2-form (\ref{bxiv}) becomes now
\bea
\omega&=&\int dx\left[\delta\log A_+\delta \left({a_+^{\prime}
\over A_+^{\prime}}-{A_+\over A_+^{\prime}}\left({a_+^{\prime}\over A_+^{\prime}}
\right)^{\prime}\right)^{\prime} \right.
\nonumber \\ 
&&\left.+\delta \log A_-\delta \left({a_-^{\prime}\over A_-^{\prime}}
-{A_-\over A_-^{\prime}}\left({a_-^{\prime}\over A_-^{\prime}}
\right)^{\prime}\right)^{\prime}+\delta f\delta\pi_f\right]+\omega_b
\>,\label{cxii}
\eea
where the boundary term $\omega_b$ is
\bea
\omega_b&=& \int d\left[ -{1\over2}\left({\delta A_+^{\prime}\over A_+^{\prime}}
-{\delta A_-^{\prime}\over A_-^{\prime}}\right)\left(\delta\left({a_+^{\prime}
\over A_+^{\prime}}\right)-\delta\left({a_-^{\prime}\over A_-^{\prime}}\right)\right)\right.\nonumber\\
&&+\delta A_+ \delta \left( {a_+^{\prime\prime}\over A_+^{\prime 2}}-
{a_+^{\prime}A_+^{\prime\prime}\over A_+^{\prime 3}}\right)
+\delta A_-\delta\left({a_-^{\prime\prime}\over A_-^{\prime 2}}
-{a_-^{\prime}A_-^{\prime\prime}\over A_-^{\prime 3}}\right)\nonumber\\
&&+{1\over2}{\lambda^2\over 1+{\lambda^2\over2}A_+A_-}\left[\left({\delta
A_+^{\prime}\over A_+^{\prime}}-{\delta A_-^{\prime}\over A_-^{\prime}}
\right)\delta\left( a_+A_--A_-a_+\right)\right.\nonumber\\
&&\left.+\delta\left({a_+^{\prime}\over A_+{\prime}}
+{a_-^{\prime}\over A_-^{\prime}}\right)\delta\left(A_+A_-\right)-
2\left(\delta a_+\delta A_-+\delta a_-\delta A_+\right)\right]\nonumber\\
&&-{1\over4}{\lambda^4\over \left(1+{\lambda^2\over2}A_+A_-\right)^2}
\left[\left(a_+A_--a_-A_+\right)\left({\delta A_+^{\prime}\over A_+^{\prime}}
-{\delta A_-^{\prime}\over A_-^{\prime}}\right)\delta\left(A_+A_-\right)\right.\nonumber\\
&&\left.\left.-2\left(a_+A_--a_-A_+\right)\delta A_+\delta A_-\right]\right]
\>.\label{cxiii}
\eea
Performing now the transformation
\be
X^{\pm}=\log A_{\pm}
\>,\label{cxiv}
\ee
\be
\Pi_{\pm}=\left({a_{\pm}^{\prime}\over A_{\pm}^{\prime}}-{A_{\pm}\over A_{\pm}^{\prime}}
\left({a_{\pm}^{\prime}\over A_{\pm}^{\prime}}\right)^{\prime}\right)^{\prime}
\>,\label{cxv}
\ee
the 2-form (\ref{cxii}) converts into (\ref{bxx}) and the constraints adopts the 
form of a parametrized field theory (\ref{bxxi}).
The composition of (\ref{cvi}-\ref{cix}) with the inverse of (\ref{cxiv}-\ref{cxv})
\be
A_{\pm}=e^{ X^{\pm}}
\>,\label{cxvi}
\ee
\be
a_{\pm}=- \int^x e^{X^{\pm}} X^{\pm\prime} \int^x e^{X^{\pm}} X^{\pm\prime}\int^x 
e^{-X^{\pm}}\Pi_{\pm}
\>,\label{cxvii}
\ee
defines, up to boundary terms, a canonical transformation.
 If we consider the case of a closed spatial section and impose 
 the condition (\ref{bxxv}) then the boundary contribution $\omega_b$
 \be
 \omega_b=-\delta \left[{a_+^{\prime}\over A_+^{\prime}}-{A_+\over A_+^{\prime}}
 \left(a_+^{\prime}\over A_+^{\prime}\right)^{\prime}-{a_-^{\prime}\over A_-^{\prime}}+
 {A_-\over A_-^{\prime}}\left({a_-^{\prime}\over A_-^{\prime}}\right)^{\prime}\right]
 \left(0\right)
 \delta \left(X^+\left(2\pi\right)-X^+\left(0\right)\right)
 \>,\label{cxviii}
 \ee 
  vanishes.
  As in the CGHS and exponential models, the additional transformation
(\ref{bxxvi}-\ref{bxxvii}) maps the theory into a bosonic string theory propagating on a 
3-dimensional Minkowskian target space.
Therefore the 
quantum analysis can be carried out along the lines of \cite{Benedict,Kuchar2} as explained
in the previous section, although now the integral conditions are
stronger
\be
\int_0^{2\pi}dx \ \Pi_{\pm}=0
\>.\label{cxix}
\ee
It follows immediately from (\ref{cxv}) and the monodromy properties of the fields
$A_{\pm},a_{\pm}$.

 The fact that a canonical transformation that 
relates the dilaton gravity to a string theory can be 
found for several 
different cases seems to indicate that there might be more models with this 
property, perhaps including the spherically symmetric Einstein gravity
 (i.e, $V \propto {1\over \sqrt{\phi}}$).
\newline
\newline
J. C. acknowledges the Generalitat Valenciana for a FPI fellowship.
D. J. N. acknowledges the Ministerio de Educaci\'{o}n y Cultura for a
FPI fellowship.
We would like to thank A. Mikovic,
M. Navarro and C. F. Talavera for useful discussions.


\begin{thebibliography}{99}
\bibitem{CGHS}
C.  G.  Callan, S.  B.  Giddings, J.  A.  Harvey and A.  Strominger, {\it Phys. Rev.} D 45 (1992) 1005.
\bibitem{Strominger}
A.  Strominger, {\it "Les Houches Lectures on Black Holes"} hep-th/9501071 (and references therein).
\newline
A.  Mikovic, {\it Phys. Lett.} B 291 (1992) 19 ; {\it Phys. Lett.} B 355 (1995) 85.
\newline
S.  Hirano, Y. Kazama and Y. Satoh, {\it Phys. Rev} D 48 (1993) 1687.
\newline
A.  Mikovic and M.  Navarro, {\it Phys. Lett.} B 315 (1993) 267.
\newline
K.  Schoutens, E.  Verlinde and H.  Verlinde, {\it Phys. Rev}. D 48 (1993) 2670.           
 \newline
 J.  Navarro-Salas,  M.  Navarro and C.  F.  Talavera, {\it Phys. Rev.} D 52
 (1995) 6831.
 \bibitem{Cangemi}
 D.  Cangemi, R.  Jackiw and B.  Zwiebach, {\it Ann. Phys} (N.Y) 245 (1996) 408.
 \bibitem{Kuchar1}
 K.  V.  Kuchar,  {\it Phys. Rev.} D 39 (1989) 1579 ; {\it Phys. Rev.} D 39 (1989) 2263.
 \newline
 K.  V.  Kuchar and C.  G. Torre, {\it J. Math. Phys.} 30 (1989) 1769. 
 \bibitem{Louis}
 D.  Louis-Martinez, J.  Gegenberg and G.  Kunstatter, {\it Phys. Lett.} B 321 (1994) 193.
 $\ \ \ $
 T.  Strobl, {\it Phys.  Rev.} D 50 (1994) 7346.
 \bibitem{Benedict}
 E.  Benedict, R.  Jackiw and H.-J.  Lee,  {\it Phys. Rev.} D 54 (1996) 6213.
 \newline
 R.  Jackiw, {\it Solutions to a Quantal Gravity-Matter Field Theory on a line},
 $2^{nd}$ Conference on Constrained Dynamics and Quantum Gravity {\it Santa
 Margherita, Italy, September 1996}, gr-qc/9612052.
 \bibitem{Kuchar2}
 K.  V.  Kuchar, J.  D.  Romano,  M.  Varadarajan, {\it Phys. Rev.} D 55 (1997) 795.
\bibitem{Louis2}
D.  Louis-Martinez, {\it Dirac Quantization of Two-Dimensional Dilaton 
Gravity Minimally Coupled to N Massless Scalar Fields}, hep-th/961103/. 
  \bibitem{Mann}
 R.  Mann,  {\it Nucl. Phys.} B 418 (1994) 231.
 \bibitem{Cruz1}
 J.  Cruz and J. Navarro-Salas,  {\it Phys. Lett.} B 387 (1996) 51.
\bibitem{Cruz2}
 J.  Cruz,  J.  Navarro-Salas, M.  Navarro and C.  F.  Talavera, {\it "Conformal and Non-Conformal
 Symmetries in 2D Dilaton Gravity"} {\it Phys. Lett.} B ( to appear).

 \bibitem{Kuchar3}
 K.  V.  Kuchar, {\it J. Math. Phys}. 17 (1976) 801.
 \bibitem{BCT}
 E.  Braaten, T.  Curtright and C.  Thorn, {\it Phys. Lett}. 118 B (1982) 115.
 
 \bibitem{HJ}
E. D'Hoker and R. Jackiw, {\it Phys. Rev}. D 26 (1982) 3517.

 \bibitem{GN}
 J.  L.  Gervais and A. Neveu, {\it Nucl. Phys} B 119 (1982) 59 ; B 209 (1982) 125
 ; B 224 (1983) 329 ; B 238 (1984) 125.
\bibitem{RJ}
R. Jackiw, {\it Geometry and Symmetry Breaking in the Liouville Theory}, in
{\it Progress in Quantum Field  Theory}, H. Ezawa and S. Kamefuchi eds.
Elsevier 1986.
 \bibitem{Jackiw}
 R.  Jackiw, in {\it "Quantum theory of gravity"}, ed. S. Christensen (Adam Hilger, Bristol, 1984)
 p.403.
$\ \ \ $
 C  Teitelboim, in {\it "Quantum theory of gravity"}, ed. S. Christensen (Adam Hilger, Bristol, 1984)
 p.327.

 \end{thebibliography}
 \end{document}